\documentclass[a4paper,twoside]{article}

\usepackage{epsfig}
\usepackage{subcaption}
\usepackage{calc}
\usepackage{amssymb}
\usepackage{amstext}
\usepackage{amsmath}
\usepackage{amsthm}
\usepackage{multicol}
\usepackage{pslatex}
\usepackage{apalike}
\usepackage{algorithm2e}
\usepackage[bottom]{footmisc}
\usepackage{url}
\usepackage{SCITEPRESS}     % Please add other packages that you may need BEFORE the SCITEPRESS.sty package.

\begin{document}

\title{Are All Genders Equal in the Eyes of Algorithms? -\\ Analysing Search and Retrieval Algorithms for Algorithmic Gender Fairness}

\author{\authorname{Stefanie Urchs\sup{1,2}\orcidAuthor{0000-0002-1118-4330
}, Veronika Thurner\sup{1}\orcidAuthor{0000-0002-9116-390X
}, Matthias Aßenmacher\sup{2,3}\orcidAuthor{0000-0003-2154-5774}, Ludwig Bothmann\sup{2}, Christian Heumann\sup{2}\orcidAuthor{0000-0002-4718-595X}, Stephanie Thiemichen\sup{1}\orcidAuthor{0009-0001-8146-9438}}
\affiliation{\sup{1}Faculty for Computer Science and Mathematics, Hochschule München University of Applied Sciences, Munich, Germany}
\affiliation{\sup{2}Department of Statistics, LMU Munich, Munich, Germany}
\affiliation{\sup{3}Munich Center for Machine Learning (MCML), LMU Munich, Munich, Germany}
\email{\{stefanie.urchs, veronika.thurner, stephanie.thiemichen\}@hm.edu, \\ \{matthias, ludwig.bothmann, christian.heumann@stat.uni-muenchen.de\}
}
}

\keywords{algorithmic fairness, academic visibility, information retrieval, search engines, gender fairness}

\abstract{Algorithmic systems such as search engines and information retrieval platforms significantly influence academic visibility and the dissemination of knowledge. Despite assumptions of neutrality, these systems can reproduce or reinforce societal biases, including those related to gender. This paper introduces and applies a bias-preserving definition of algorithmic gender fairness, which assesses whether algorithmic outputs reflect real-world gender distributions without introducing or amplifying disparities. Using a heterogeneous dataset of academic profiles from German universities and universities of applied sciences, we analyse gender differences in metadata completeness, publication retrieval in academic databases, and visibility in Google search results. While we observe no overt algorithmic discrimination, our findings reveal subtle but consistent imbalances: male professors are associated with a greater number of search results and more aligned publication records, while female professors display higher variability in digital visibility. These patterns reflect the interplay between platform algorithms, institutional curation, and individual self-presentation. Our study highlights the need for fairness evaluations that account for both technical performance and representational equality in digital systems.}

\onecolumn \maketitle \normalsize \setcounter{footnote}{0} \vfill

\section{\uppercase{Introduction}}
Algorithms are increasingly embedded in nearly every aspect of our daily lives, shaping the information we encounter and influencing our perceptions and decisions. From social media recommendations to online shopping suggestions, algorithmic processes impact what we see, how we engage, and ultimately how we make choices. Among these, algorithms in search engines and publication databases have significant power in determining which information, content, and experts are made visible to users, directly influencing public knowledge, career opportunities, and academic visibility. For instance, studies have shown that job advertisements displayed by search engines can be targeted by gender~\cite{datta2015automated,63f4b34397794f5b8699dd81ca3e628a}, image search results prefer white individuals~\cite{10.1007/978-3-030-78818-6_5} and text-based search results sexualise woman, especially from the global south~\cite{Urman_2022}, raising significant concerns about the presence and impact of gender-based bias in these systems. Such examples underscore the urgency of examining and defining algorithmic fairness, particularly regarding gender representation, as these biases risk perpetuating and amplifying existing societal inequities.

Algorithmic gender fairness is essential because these biases are not merely technical flaws but reflections of deeper societal structures embedded in data and system design. Algorithms do not operate in isolation; they are shaped by the data they are trained on, the objectives they are optimised for, and the societal context in which they function. Addressing gender fairness requires navigating the intersection of mathematical criteria and social implications, as technical fixes alone cannot resolve biases rooted in historical and structural inequalities. Without a well-defined framework for fairness, efforts to mitigate algorithmic discrimination risk being inconsistent or even counterproductive. Therefore, a clear and robust definition of algorithmic gender fairness is crucial, not only to prevent direct and indirect discrimination but also to establish transparency, accountability, and trust in automated systems. 

Building upon existing research in algorithmic fairness and algorithmic gender fairness, this work contributes to the ongoing discourse by proposing and empirically testing a definition of algorithmic gender fairness. While many studies have explored fairness in algorithms, our approach focuses on evaluating two influential types of systems: publication database retrieval algorithms and Google's search engine. These algorithms play a crucial role in shaping public visibility and access to information, making them particularly impactful subjects for analysis. By applying our fairness definition to these systems, we aim to offer insights into their performance, identify improvement areas, and contribute to developing more transparent, accountable, and inclusive algorithmic designs.

\section{\uppercase{Background}}
To define algorithmic gender fairness, we begin by outlining how we understand the core concepts of gender and fairness. Given the interdisciplinary nature of this work, the section is deliberately extensive. In the final part of the section, we first introduce how information retrieval and search engines work in general, providing the necessary technical background for readers unfamiliar with the field. We then review existing research on algorithmic fairness in these domains and highlight how our approach differs from previous work.

\subsection{Gender}
The term ``gender'' encompasses at least three distinct concepts: linguistic gender, sex, and social gender. Each concept has unique implications in various professional and private contexts, especially when considering algorithmic representation, identity, and fairness issues. Linguistic or grammatical gender is defined as ``\textit{[...] grammatical gender in the narrow sense, which involves a more or less explicit correlation between nominal classes and biological gender (sex).}''~\cite{janhunen2000grammatical}. In many languages, nouns and pronouns are assigned a gender, classified as feminine, masculine, or neutral, often loosely correlated with perceived biological characteristics~\cite{Kramer_2020}. This linguistic categorisation can affect the way gender roles and identities are understood culturally, as language shapes and reinforces social expectations~\cite{Konishi_1993,phillips2013can}. 

``Sex'', on the other hand, is traditionally understood as a biological categorisation, regarded as ``\textit{binary, immutable and physiological}''~\cite{10.1145/3274357}. However, a strict binary framework is increasingly recognised as insufficient for representing the full spectrum of human diversity. Intersex individuals, who may not fit the conventional definitions of feminine or masculine due to variations in physiological characteristics~\cite{Carpenter01042021}, and transgender individuals, whose gender identity differs from their sex assigned at birth~\cite{beemyn2011lives}, exemplify the limitations of this binary, immutable perspective. The presence of these identities challenges the conventional definitions of sex.

In our work, we embrace the concept of social gender, which goes beyond biological and linguistic classifications to encompass a socially constructed identity shaped by behaviours, expressions, and self-presentation. Social gender is fluid, non-binary, and co-constructed through social interactions, allowing it to evolve over time in alignment with an individual's sense of self. This perspective aligns with research that views gender not as an inherent or static characteristic but as a performative act shaped by personal expression and social context~\cite{doi:10.1177/0891243287001002002,10.1145/3531146.3534627}.

Although we adopt this inclusive understanding of gender, our study faces limitations due to the constraints in our data. The available information only allows for analysing participants within the binary gender spectrum, and we were thus unable to identify trans or intersex individuals in the dataset. As a result, our empirical analysis focuses on binary gender categories. However, the underlying framework of our proposed definition of algorithmic gender fairness remains rooted in the concept of social gender -- emphasising its non-binary, flexible, and socially co-constructed nature. We aim to contribute to a broader, more inclusive understanding of gender fairness in algorithmic systems, even as we acknowledge the current limitations of our dataset.

\subsection{Fairness}
\label{fairness}
The term ``fairness'' is increasingly used in the field of algorithmic decision-making and ``fairness-aware machine learning'' [fairML, surveys can be found, e.g., in ~\cite{caton_fairness_2024,verma_fairness_2018}]. However, few contributions concretely define the meaning of this term as a philosophical concept, with positive exceptions to be found in \cite{bothmann_what_2024,loi_is_2022,kong_are_2022}. Fairness is usually described by synonyms such as equality, justice, or the absence of bias or discrimination. 

A crucial component of fairness as a philosophical concept is that it concerns the treatment of individuals \cite{aristotle_nicomachean_2009,cambridge_dictionary_fairness_2022,dator_chapter_2017,kleinberg_inherent_2017}. The basic structure of the concept can be traced back to Aristotle and relates fairness to equality: A decision or treatment is fair if equals are treated equally and unequals are treated unequally. As \cite{bothmann_what_2024} point out, this requires the normative definition of task-specific equality, that is: Two individuals may be equal in one task (e.g., buying a croissant in a bakery), but unequal in another task (e.g., paying taxes). Deciding how to treat unequals is also a normative task.

The role of protected attributes such as gender or race is that they can normatively alter the definition of task-specific equality. For example, a society may decide that the grievance of the gender pay gap is not the responsibility of an individual and that in deciding whether to grant a loan, income should therefore be fictitiously corrected for this real-world bias; \cite{bothmann_what_2024} call this a fictitious, normatively desired (FiND) world, and advocate making decisions using data from this world rather than real-world data. \cite{wachter_bias_2021} describe such an approach as ``bias-transforming'', aiming at ``substantive equality'', because a real-world bias should be ``actively eroded'' to make the world fairer.

In contrast, \cite{wachter_bias_2021} describes approaches as ``bias-preserving'', aiming at ``formal equality'', if they try to reflect the real world as accurately as possible, i.e., without introducing new biases that may even increase the real-world biases. Many fairML metrics, such as equalised odds or predictive parity, can be categorised as bias-preserving because they measure against real-world labels but try to balance the errors thus measured across levels of the protected attribute. Sometimes the concept of bias-transforming methods is referred to as aiming for ``equity'', while bias-preserving approaches are referred to as aiming for ``equality''. In our work, we will follow a bias-preserving approach to adequately or ``correctly'' reflect individuals in the real world while prohibiting the introduction of gender bias by information retrieval algorithms or search engines (in addition to the already existing gender bias in the real world).

\subsection{Information Retrieval and Search Engines}
\label{IR_search}
Information Retrieval (IR) focuses on finding relevant material, typically text documents, to satisfy a user's information need. An information need represents the user's underlying intention or goal when seeking information. At the same time, a query explicitly represents this need, usually entered as keywords or phrases in a search engine. These concepts are fundamental in bridging the gap between human intentions and computational processing, ensuring that search systems accurately interpret and address user needs~\cite{schutze2008introduction}.

An Information Retrieval System (IRS) is a software system that efficiently stores, manages, and retrieves information from large datasets. An IRS relies on indexing and searching algorithms to match user queries with relevant documents. Retrieval systems can be categorised based on their retrieval models, with the two primary examples being Boolean Retrieval and Vector Space Retrieval. Boolean Retrieval allows users to formulate queries using logical operators such as AND, OR, and NOT, ensuring that documents are returned only if they satisfy the Boolean expression. On the other hand, the Vector Space Model represents documents and queries as vectors in a multi-dimensional space, using similarity measures like cosine similarity to rank results by relevance. In document retrieval, user queries are matched against different parts of documents, such as title, keywords, author name(s), and abstract. These metadata fields often provide valuable signals for relevance, enabling the system to prioritise results more effectively. An IRS typically employs inverted indexes, which map each term to a list of documents containing it, facilitating rapid query processing. Additionally, ranking algorithms ensure that results are retrieved and presented in an order reflecting their relevance to the user's query~\cite{schutze2008introduction}.

Recent research highlights a critical issue within IRS: the presence of biases in their structure and outcomes~\cite{Fang2022}. These biases can emerge from relevance judgment datasets, neural representations, and query formulation. Relevance judgment datasets, often regarded as gold-standard benchmarks, may carry stereotypical gender biases, propagating into ranking algorithms when IRS are trained on such datasets~\cite{10.1145/3477495.3532680}. Additionally, neural embeddings used for query and document representations, pre-trained on large corpora, are susceptible to inheriting societal biases present in those datasets~\cite{bolukbasi2016man}. Retrieval methods, especially those using neural architectures, have shown a tendency to intensify pre-existing gender biases~\cite{francazi2024initial}. Bias-aware ranking strategies, such as adversarial loss functions, bias-aware negative sampling, and query reformulation techniques (e.g., AdvBERT), have been proposed to reduce these biases while maintaining retrieval effectiveness. Researchers emphasise the importance of balancing retrieval performance with fairness, advocating for systematic evaluation metrics and datasets explicitly designed for measuring and mitigating gender biases in IRS~\cite{10.1145/3477495.3532680}. Prior work on fairness in information retrieval has largely focused on technical interventions in ranking systems (e.g., \cite{singh2018fairness}; \cite{geyik2019fairness}) or on consumer-side fairness~\cite{Ekstrand_2022}, typically evaluating search and recommendation systems in general-purpose digital platforms. In contrast, few empirical studies have investigated gender fairness in academic retrieval contexts. Our work bridges this gap by conducting a fairness audit of academic visibility, applying a bias-preserving fairness perspective to both domain-specific publication databases and general-purpose search engines. In doing so, we extend the methodological orientation of studies like \cite{10.1145/3477495.3532680} and \cite{Fang2022} to a new sociotechnical domain. For instance, Singh and Joachims~\cite{singh2018fairness} propose formal fairness constraints on exposure in rankings, ensuring that protected groups receive visibility proportional to their relevance. Their framework relies on probabilistic rankings to balance user utility and provider fairness in expectation.

Search engines are advanced Information Retrieval Systems tailored for web-scale datasets. They consist of three primary components: crawling, indexing, and query processing. Crawlers systematically fetch web pages indexed using data structures like inverted indexes. Query processing involves parsing the user's input and matching it with indexed documents. The PageRank algorithm, introduced by Google, revolutionised web search by considering the hyperlink structure of the web. Each webpage is assigned a numerical score based on the quantity and quality of incoming links. The algorithm models a ``random surfer'' who follows hyperlinks or randomly jumps to other pages. This behaviour is mathematically represented using Markov Chains, and steady-state probabilities are computed iteratively to determine the importance of each page. Search engines blend PageRank with other ranking factors, including content relevance, term proximity, and user-specific data, creating a hybrid scoring system that delivers highly accurate search results~\cite{schutze2008introduction}.

However, search engines are not immune to biases. Biases in search engines can emerge from data sources, crawling strategies, and ranking algorithms, resulting in the reinforcement of stereotypes, underrepresentation of marginalised groups, or discriminatory exposure of content. Biases may also be amplified over time through dynamic adaptation mechanisms, where user interactions create feedback loops that reinforce pre-existing biases. Addressing these biases requires mitigation strategies such as bias-aware re-ranking algorithms, adversarial training, and query reformulation techniques~\cite{Ekstrand_2022}.

Additionally, fairness concerns in search engines align with consumer fairness (ensuring users receive equally relevant and satisfying results across diverse groups) and provider fairness (ensuring content creators or document providers receive equitable exposure in rankings). Evaluation methodologies play a key role in addressing these concerns, often combining relevance metrics with fairness-aware metrics to strike a balance between accuracy and equity~\cite{Ekstrand_2022}. In industrial applications, \cite{geyik2019fairness} present a fairness-aware re-ranking framework deployed at scale in LinkedIn Talent Search. Their system enforces minimum representation thresholds through post-processing algorithms, demonstrating that fairness and utility can coexist in production systems. However, their approach is grounded in fairness-transforming principles such as demographic parity.

In practice, search engines represent a complex interplay between technical architecture, algorithmic fairness, and societal values. Continuous research and refinement are essential to ensure these systems meet efficiency and fairness criteria simultaneously~\cite{Ekstrand_2022}.

\section{\uppercase{Algorithmic Gender Fairness}}
\label{sec:AGF}
To define algorithmic gender fairness, we build upon the theoretical framework presented in Section ``Fairness" and the practical insights discussed in Section ``Information Retrieval and Search Engines". Our approach adopts a bias-preserving perspective, aiming to reflect real-world distributions without introducing new distortions or exacerbating existing gender biases.

Bias in algorithmic systems can arise from several sources, including biased training datasets, pre-existing societal inequalities, and the interaction between users and algorithmic feedback loops~\cite{european_commission2021}. Gender biases, in particular, are often perpetuated through historical inequalities encoded in data, proxies that stand in for protected attributes, and opaque decision-making processes inherent to many machine-learning systems.

At the data stage, biases can emerge from training datasets that reflect societal inequalities, including historical gender pay gaps or occupational stereotypes. These biases are often amplified when algorithms learn patterns from these datasets without critical oversight. From a bias-preserving perspective, systems should strive to reflect gender distributions accurately without further entrenching societal disparities. However, achieving this requires ongoing monitoring and transparency to detect and address unintended distortions.

At the algorithmic stage, gender biases can manifest in ranking systems, recommendation algorithms, or classification processes. Proxy variables, such as zip codes, browsing behaviour, or inferred demographic data, often serve as indirect markers for gender, leading to indirect discrimination. Mitigating these biases involves identifying such proxies and adjusting algorithmic models to ensure they do not disproportionately disadvantage individuals based on gender~\cite{european_commission2021}.

From a bias-transforming perspective, algorithms may be adjusted proactively to counteract historical inequalities and actively reshape outcomes. Such approaches aim for substantive equality, where systems not only avoid perpetuating existing biases but actively correct for them by introducing calibrated adjustments to outputs~\cite{european_commission2021}. Such fairness interventions are often formalised as constrained optimisation problems, where utility (e.g., accuracy or public safety) is maximised subject to fairness constraints. \cite{corbett2017algorithmic} demonstrate that implementing common fairness definitions, such as statistical parity or predictive equality, typically requires group-specific decision thresholds, a trade-off that can reduce utility or violate principles of equal treatment.

Transparency and explainability remain central challenges in algorithmic gender fairness. The opacity of many systems, particularly those based on deep-learning architectures, makes it difficult to detect and address gender biases effectively. Without clear explanations of how decisions are reached, it becomes challenging to hold systems accountable for gender-discriminatory outcomes.

Additionally, intersectionality plays a crucial role in algorithmic gender fairness. Gender does not exist in isolation but intersects with other protected attributes such as race, age, or socio-economic status, leading to compounded forms of bias and discrimination. Addressing intersectionality requires fairness-aware metrics that account for these overlapping dimensions~\cite{european_commission2021}.

Our approach focuses on bias-preserving fairness as the guiding principle, ensuring that algorithmic systems in information retrieval and search engines reflect real-world gender distributions without introducing additional biases. While our approach focuses on preserving bias patterns as they exist in real-world data, many prior works have proposed alternative fairness frameworks. \"Zliobait\.e~\cite{zliobaite2017measuring} offers a systematic overview of such fairness definitions in algorithmic decision-making, highlighting group fairness notions such as statistical parity, conditional parity, and predictive parity, as well as individual fairness principles based on similarity of treatment. While bias-transforming approaches, which aim to correct historical inequalities proactively, offer an appealing vision of fairness, they require defining an ideal dataset or outcome, a ``perfect world'', to serve as a benchmark. However, defining such an ideal world is inherently challenging, given the vast diversity of cultural, social, and political value systems across the globe. Even if we attempted to define it, measuring an ideal world would remain an insurmountable task, as no dataset could comprehensively capture such a reality.

Given these constraints, we adopt a bias-preserving approach, which evaluates whether algorithms accurately reflect and replicate the analogue reality within the digital domain without amplifying existing biases. This approach leverages measurable real-world data, allowing us to assess algorithmic outcomes in relation to observed societal distributions. 

Therefore, we define algorithmic gender fairness as:
\begin{quote}
    \textit{The ability of algorithmic systems, particularly in information retrieval and search engines, to accurately reflect real-world gender distributions and representations in their outputs without introducing, amplifying, or reinforcing existing biases.} 
\end{quote}

In this paper, we apply the above definition of algorithmic gender fairness to evaluate real-world systems that mediate academic visibility. While prior studies have primarily focused on technical fairness interventions or theoretical proposals, our contribution lies in conducting a fairness audit grounded in this definition, using empirical data from both domain-specific academic databases and a general-purpose search engine. By doing so, we extend the application of fairness frameworks to a previously underexplored domain: the digital representation of academic expertise.

\section{\uppercase{Experiments}}
We test our notion of algorithmic gender fairness by analysing the online visibility of professors through two distinct types of algorithmic systems: search algorithms, exemplified by Google, and information retrieval algorithms used in academic publication databases. While Google clearly ranks results through its proprietary search algorithm, the publication databases return results based on "relevance", a criterion that remains undefined by the platforms. Consequently, we do not compare the results directly but instead analyse each system separately to explore how algorithmic structures may influence visibility across gender lines.

\subsection{Data}
\label{subsec:data}
The data for this study stems from a broader research project that investigated the visibility of female professors at universities of applied sciences (UAS)\footnote{UAS are a distinct feature of the German higher education system. They focus on practice-oriented teaching and maintain close ties to industry. Compared to traditional universities, they generally have smaller student groups and place less emphasis on theoretical research. Within the German academic system, traditional universities often view UAS as less prestigious due to their more applied, less theory-driven orientation.} in Germany. The full dataset includes professors from different institutional types (universities and universities of applied sciences) and academic disciplines (computer science and social work/social pedagogy). This heterogeneity was intentional: to capture a broad spectrum of academic visibility, we aimed for maximum variation within the German academic landscape. Including both institutional types reflects structural differences in prestige, mission, and digital presence. Moreover, computer science and social sciences follow distinct publication cultures: computer science is predominantly conference-driven, while social scientists typically publish in journals.

As the main focus of the project lay on female professors at UAS, we manually collected the full population of women professors working in the departments of computer science and social work at these institutions. To provide a meaningful comparison, we additionally included random samples of male professors at UAS, as well as female and male professors from traditional universities in comparable fields. For university-level social science, we focused on social pedagogy, as the field of social work is not formally established at universities. The comparison samples were drawn from all German UAS and universities that host relevant departments in the selected disciplines. Table~\ref{tab:sample} summarises the resulting sample sizes for both the full dataset and the balanced subsample used in downstream analyses.

For the Google-based analysis, we used the full dataset. For the publication database analysis, we drew on a balanced subsample of 80 professors (40 female, 40 male), randomly selected to ensure equal representation across institutional types. We relied on a subsample of the full dataset because extracting publication lists required manual effort. Since each professor curated their own list individually and in non-standardised formats, the extraction process could not be automated.

Gender was inferred from the presentation on university profiles and treated as binary due to the limitations of available data. Public websites typically included names and profile pictures only, so gender was manually inferred based on these attributes. We acknowledge that this is not best practice, as it does not allow individuals to self-identify. However, contacting each professor individually was not feasible. The student responsible for data collection was instructed to assign a gender only when absolutely certain; otherwise, entries were to be marked as \textit{unknown}. In practice, no such cases occurred.

\begin{table}[h!]
    \centering
    \caption{Sample size of professors of the full data set and the subsample. The full dataset contains all female professors at UAS in the departments of computer science and social work. For all other categories, a random sample of 50 professors was used. The random sample was used as a comparison group for the main focus of the project, female professors at UAS.}
    \begin{tabular}{p{3.5cm}|p{1.5cm}|p{1.5cm}}
        & Full Dataset & Subsample \\ \hline
         Female Professors UAS\newline Computer Science & 219 & 10\\
         Female Professors UAS\newline Social Work & 863 & 10\\
         Female Professors Uni\newline Computer Science & 50 & 10\\
         Female Professors Uni\newline Social Pedagogy & 50 & 10\\
         Male Professors UAS Computer\newline Science & 50 & 10\\
         Male Professors UAS\newline Social Work & 50 & 10\\
         Male Professors Uni\newline Computer Science & 50 & 10\\
         Male Professors Uni\newline Social Pedagogy & 50 & 10\\
    \end{tabular}
    \label{tab:sample}
\end{table}

For each professor, the following information was collected:
\begin{itemize}
    \item Name and title
    \item Gender (inferred)
    \item Institutional affiliation
    \item Reported keywords
    \item Presence of a CV and/or picture on the university profile
    \item Publication list on the university profile
\end{itemize}

Because professors manage their own profiles and present publication lists in diverse formats, all data was manually extracted.

\subsection{Experimental Design}
\label{design}
To examine gendered visibility in digital environments, we analyse three interconnected layers of representation: Google search results, academic publication databases, and university profiles. Each serves a distinct role in how professors are made visible, discovered, and contextualised online.

\textbf{Google Search Results.}
We began our analysis by examining broader forms of digital visibility through Google Search. For each professor in the full sample, we conducted a name-based search that included their institutional affiliation and collected the first 100 search results. These results were categorised into the following types:
\begin{itemize}
    \item university
    \item social media
    \item research institutes
    \item newspapers/media
    \item research profiles
    \item publication databases/preprint servers
\end{itemize}
We analysed the number, type, and ranking position of these results to identify gendered patterns in digital visibility, with a particular focus on whether algorithmic search systems shape differential representations of female and male professors. Given the broader reach of search engines and the structured nature of Google results, this part of the analysis serves as the primary basis for evaluating algorithmic gender fairness in our study.

\textbf{Publication Databases.}
To complement the Google-based visibility analysis, we also examined how academic content is retrieved in publication databases, a step that reflects common search strategies used by science journalists and other knowledge intermediaries. It is a typical workflow to begin by querying databases for topic-relevant keywords, and only after identifying promising names, turn to search engines like Google for more context. To honour this process, we conducted an additional exploratory analysis based on academic keyword searches.

For this analysis, we focused on a balanced subsample of 80 professors. From their university profiles, we compiled all self-reported keywords and queried them individually in three major academic databases:  the ACM Digital Library\footnote{\url{https://dl.acm.org/}} (used for publication in computer science), Springer Link\footnote{\url{https://link.springer.com/}} (used for publication in computer science and social sciences), and Beltz\footnote{\url{https://www.beltz.de/}} used for publication in social sciences). For each professor in our subsample, we extracted all self-reported keywords from their university profiles and compiled them into a single list. Each keyword in the list was queried individually in the respective databases, and for each query, we collected the top 1,000 results. We then attempted to match retrieved publications to professors based on their names. We attempted to match retrieved publications to professors based on their names, using either the full first and last name or the first initial and last name. Given the limited available information, this was the most feasible matching strategy, despite the potential for false positives. However, a manual review of the matches confirmed that they appeared valid.

Because publication lists were not uniformly available for all individuals in the full dataset, this analysis was limited to a balanced subsample of 80 professors. While this sample size does not support generalisable claims, it provides initial insights into how academic content is retrieved and associated with named individuals in these databases. The results should be interpreted with caution, particularly as the databases do not disclose how their ranking is determined; search results are typically ordered by ``relevance,” but the underlying criteria remain opaque. As a result, this part of the analysis serves primarily as an exploratory context. However, following our definition of algorithmic gender fairness introduced in Section ``Algorithmic Gender Fairness", we use the gender composition of this subsample as a reference for the real-world distribution against which retrieval outputs are compared.

\textbf{University Profile Completeness.}
In addition to Google search results and publication databases, we analysed the content of university profiles to capture how professors are presented on their institutional websites. As detailed in Subsection ``Data", this information was manually extracted and includes the presence of a CV, a profile picture, and a publication list. These profiles represent structured, publicly accessible data curated by the professors themselves or their institutions. In line with our definition of algorithmic gender fairness, we treat them as a form of real-world data that serves as a reference point for evaluating how academic professionals are represented in digital environments such as search engines.

\subsection{Findings}
This section presents the main findings from our analysis, structured across three areas: Google search results, academic publication databases and the completeness of university profiles.

\textbf{Publication Databases.} Across all keyword-based database queries, we retrieved a total of 48,541 unique publications. However, only 44 of these could be matched to professors in our subsample, using either their full name or first initial and surname. This surprisingly low match rate highlights a significant disconnect between the academic work professors report and what is discoverable through our keyword-based database searches.

Several factors likely contribute to this outcome. Most importantly, our queries were limited to three specific publication outlets, the ACM Digital Library, Springer Link, and Beltz, chosen because they allow for automated querying and due to their relevance in informatics and social sciences. As a result, publications in other venues were not included. In addition, professors may not have published under the exact keywords they listed on their university profiles, or the terms may have been too broad or too specific to yield meaningful matches. Keyword searches may also miss publications where the terms are not prominent in titles or abstracts. Further limitations stem from the databases themselves: relevance-based ranking may exclude pertinent results, and name matching can lead to both, false negatives and false positives. If multiple individuals share the same name, our approach may have incorrectly assigned a publication to a professor in the sample.

\begin{figure}[h]
    \centering
    \includegraphics[width=0.45\textwidth]{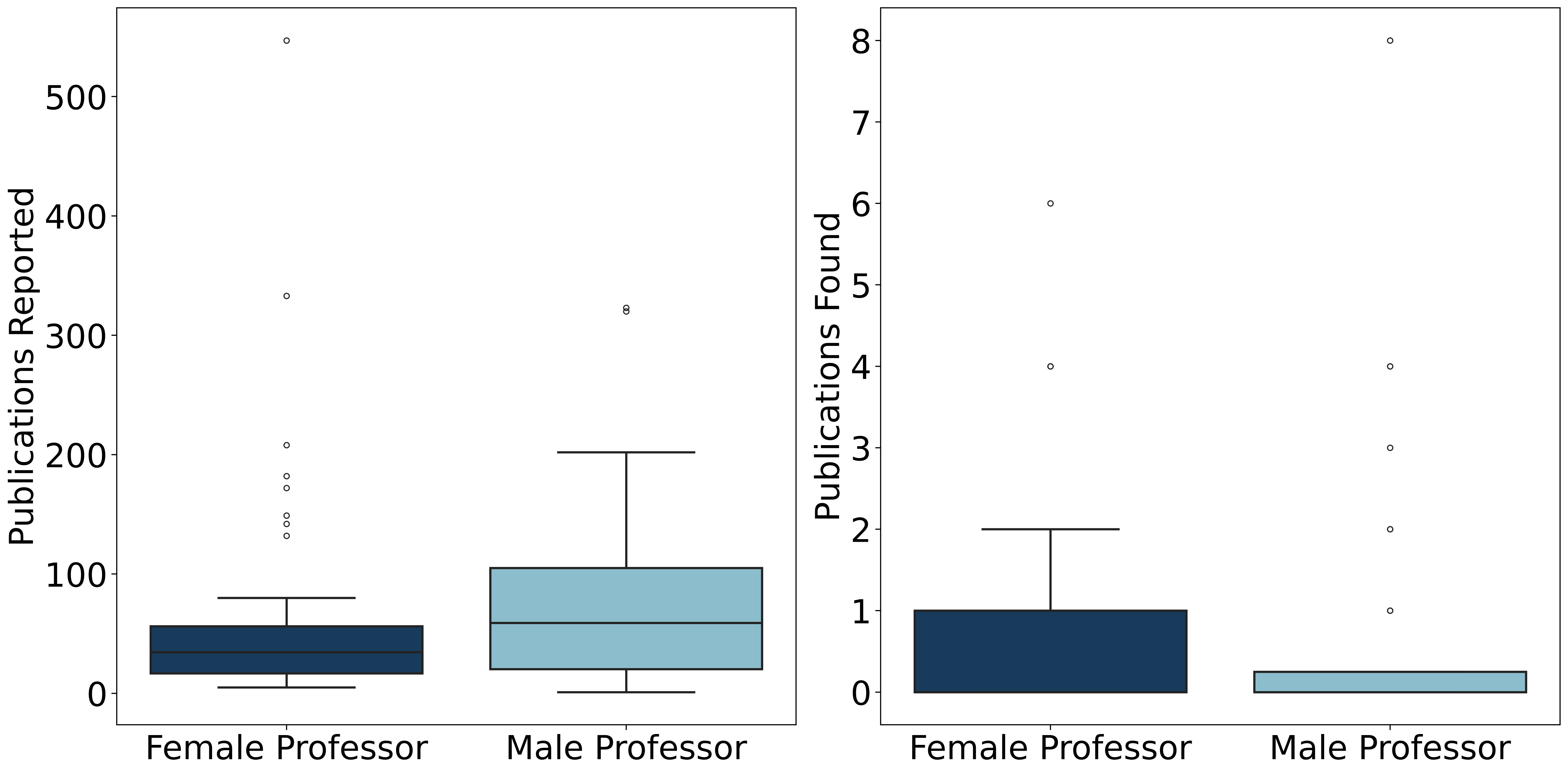}
    \caption{Self-reported versus found publications (via keywords), per person.}
    \label{fig:report_vs_found}
\end{figure}

Figure~\ref{fig:report_vs_found} shows that male professors generally reported more publications, including several extreme outliers. However, very few publications were actually found through database searches for either gender, highlighting the limited recall of keyword-based retrieval in this context.

\begin{figure}[h]
    \centering
    \includegraphics[width=0.45\textwidth]{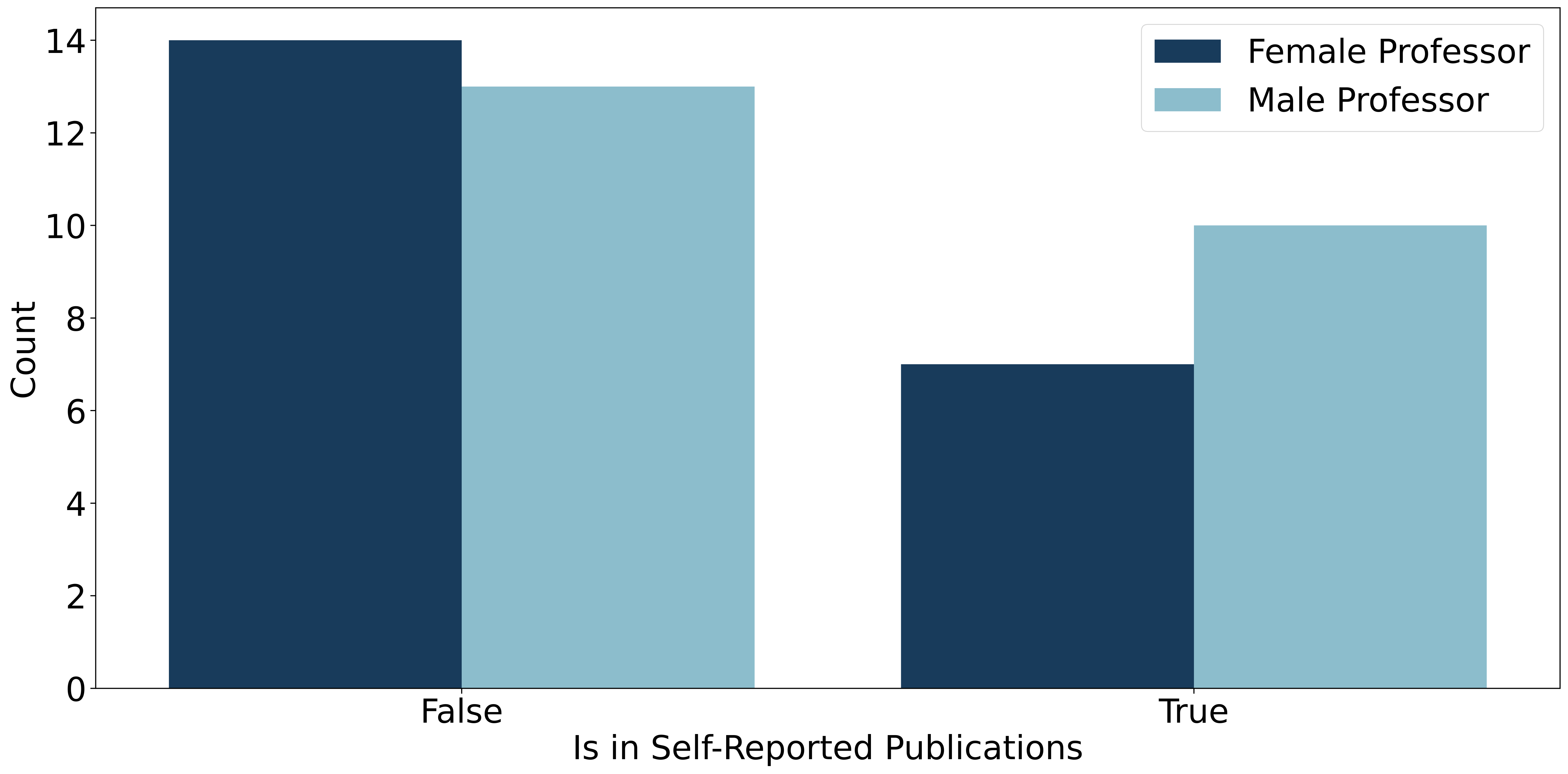}
    \caption{Publications retrieved from databases (via keywords) that also appeared in self-reported lists.}
    \label{fig:found_in_report}
\end{figure}

Figure~\ref{fig:found_in_report} shows how many of the matched publications were also part of the professors’ self-reported publication lists. While female professors had a slightly higher number of matches, the majority of retrieved publications were not part of the self-reported lists for either group. This again suggests that keyword selection and platform coverage substantially shape which publications become visible through database queries.

\begin{figure}[h]
    \centering
    \includegraphics[width=0.45\textwidth]{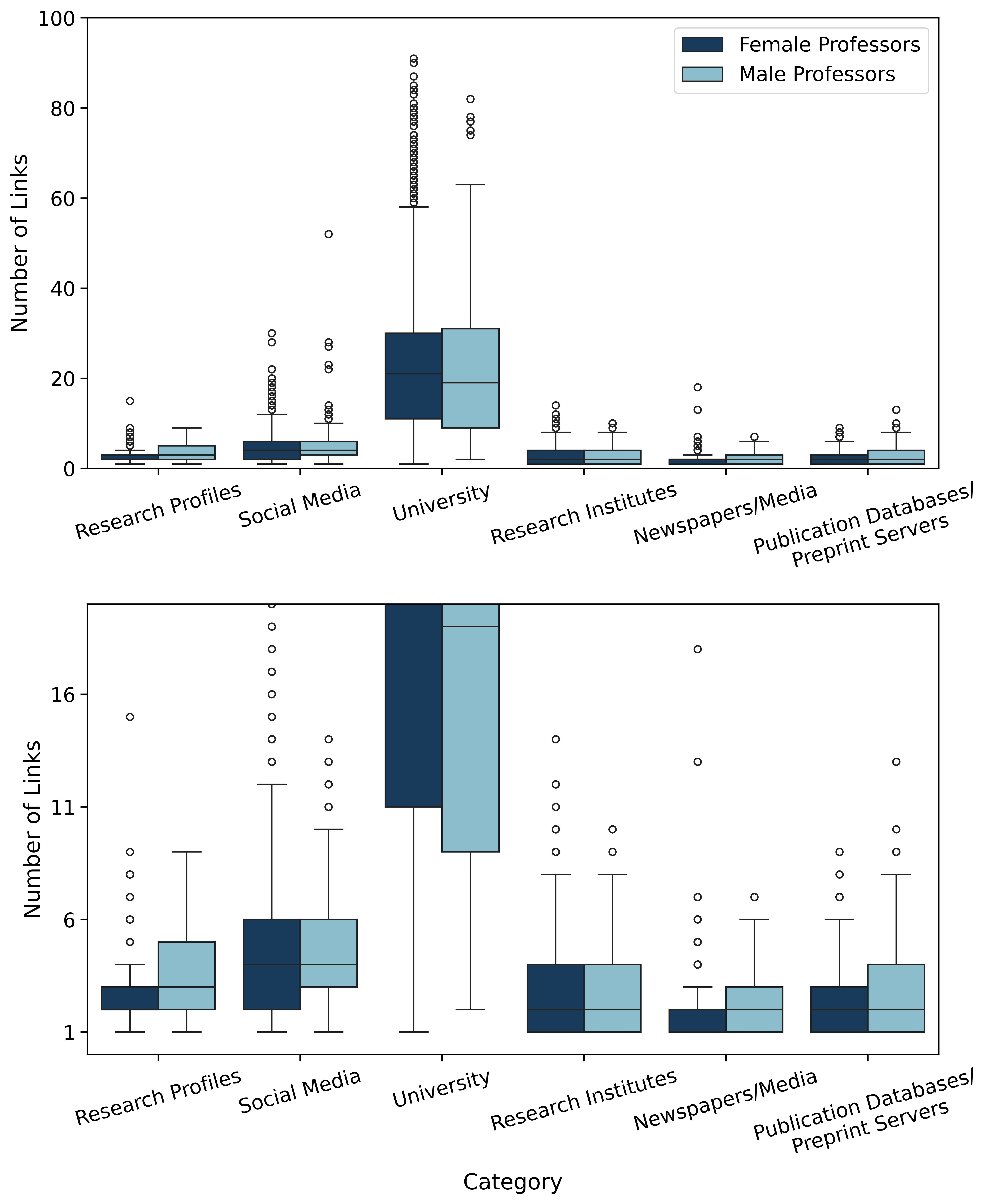}
    \caption{Number of links per category for female and male professors. The top plot shows full data; the bottom plot zooms into low-frequency categories.}
    \label{fig:link_counts}
\end{figure}

\textbf{Google Results.}
We next examined how professors are represented across broader digital platforms using Google search results. For each professor in the full sample, we retrieved and categorised the first 100 results. Figure~\ref{fig:link_counts} shows the number of links per category, grouped by gender. University-related links were the most common for both female and male professors. Overall, male professors had more links, with a higher median and more variation. Female professors showed a tendency for outliers and more individuals having few or very few links.

\begin{figure}[h]
    \centering
    \includegraphics[width=0.45\textwidth]{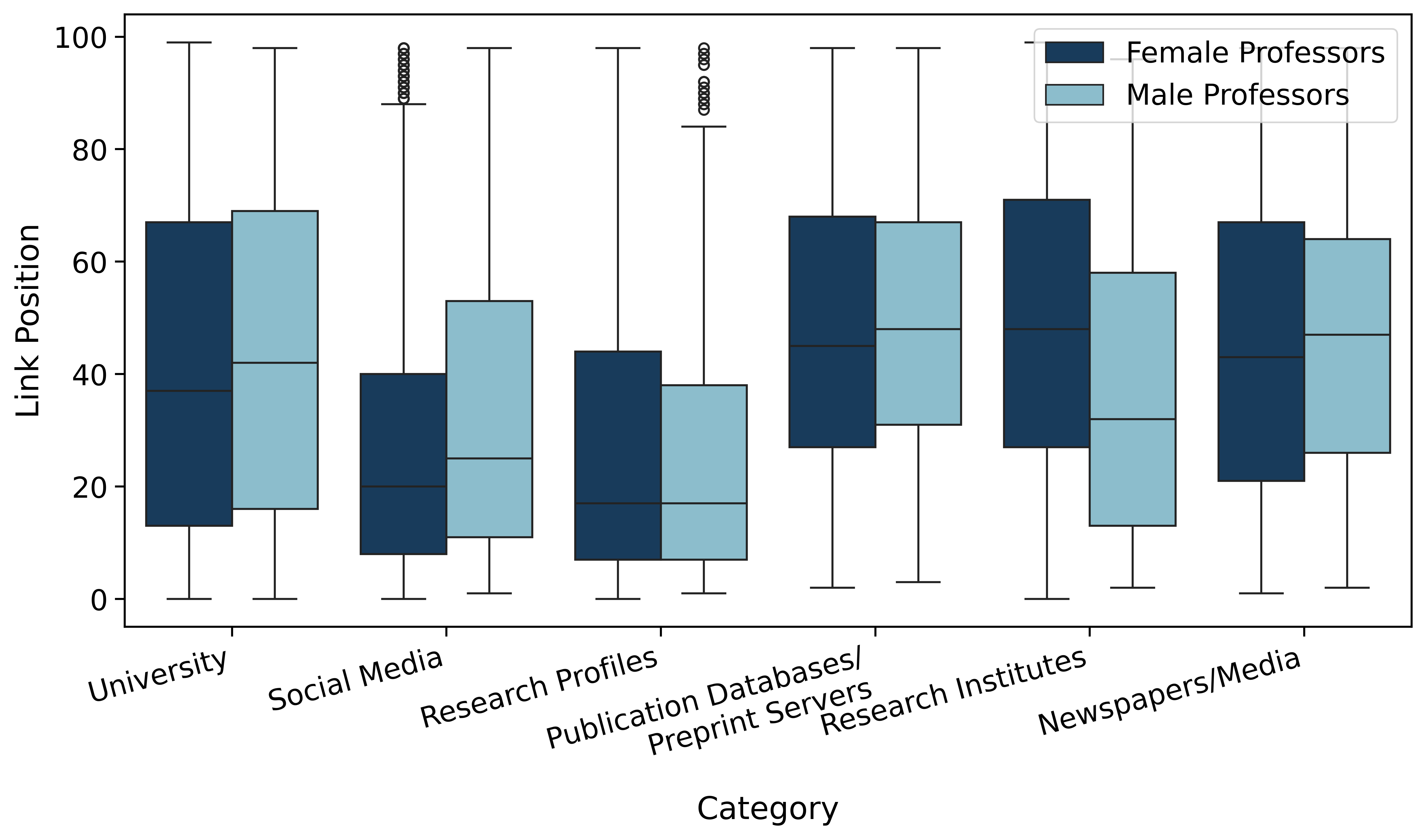}
    \caption{Ranking position of Google search results across categories, by gender. Lower values indicate higher ranking.}
    \label{fig:link_position}
\end{figure}

Figure~\ref{fig:link_position} presents the ranking positions of these links. Female professors' university links tended to appear slightly higher in the result lists, while male professors had better visibility in categories like research profiles and social media. Although the differences are subtle, they contribute to an overall pattern of gendered variation in search engine visibility.

\textbf{University Profile Completeness.}
We also examined the content of university profiles for all professors in the full dataset. As shown in Table~\ref{tab:cv_pic}, most professors included both a CV and a profile picture, and over two-thirds also provided a publication list. Female professors were slightly more likely to include a CV and a publication list, while male professors were marginally more likely to include a picture. 

\begin{table}[h]
    \centering
    \caption{University profile completeness for the full dataset: CV, picture and publication list inclusion. The numbers should be interpreted as a percentage of female professors or a percentage of male professors, depending on the line. Therefore, rows do not add up to 100\%.}
    \begin{tabular}{l|c|c|p{1.5cm}}
        & CV & Picture & Publication Lists\\\hline
        Female Professor & 68.9\% & 85.0\% & 70.2\% \\
        Male Professor & 62.5\% & 85.9\% & 66.2\%
    \end{tabular}
    \label{tab:cv_pic}
\end{table}

\subsection{Discussion}
Our findings indicate that digital visibility in academic contexts is subtly but consistently gendered. This becomes particularly evident when analysing how algorithmic systems represent female and male professors across different platforms. While we did not observe overt algorithmic discrimination, patterns in both database retrieval and Google search results suggest that gender affects how academic expertise is surfaced and made visible.

In publication databases, we found a substantial gap between self-reported and retrieved publications. This gap stems from multiple sources: limited platform coverage (restricted to three specific outlets), reliance on self-reported keywords that often did not align with actual publication metadata, and opaque ``relevance”-based ranking mechanisms that are not designed to ensure fair or comprehensive representation. Additionally, name-based matching introduces ambiguity, especially for common names. Although the sample was too small to draw generalisable conclusions, male professors showed slightly higher match rates, pointing to possible gendered differences in how academic outputs are indexed and surfaced.

In contrast, our analysis of Google search results, conducted on the full dataset, revealed clearer patterns. Male professors were consistently associated with a higher number of links across most categories. However, the distributions were not uniformly more concentrated for male professors. While they had higher medians in several categories, the spread of results varied by category and was not consistently narrower than that of female professors. Female professors showed greater variability overall, with more frequent low-end outliers, particularly in categories such as university and research profiles. When considering the ranking of results, female professors’ links tended to appear slightly higher in several categories, including social media, research profiles, and publication databases. In other categories, such as university and newspapers/media, the ranking distributions were largely comparable across genders. These findings suggest that while female professors are not disadvantaged in terms of ranking within categories, the lower number of links may still reduce their overall discoverability in search results.

A possible factor contributing to the lower number of search results for female professors is the way academic profiles are structured on institutional websites. While profile completeness was generally high across the sample, we observed small gender differences: female professors were slightly more likely to include CVs and publication lists, whereas male professors more frequently provided a profile picture. Since images and structured information (such as publication entries or CVs) can be indexed differently by search engines, these differences in self-presentation may influence how easily professors are linked to relevant content. What is particularly striking, however, is that despite female professors providing slightly more structured academic information on their university profiles, they were less visible in several key categories of Google search results, most notably ``research profiles,” ``publication databases/preprint servers,” ``newspapers/media,” and ``university.” In other words, even though they appear to invest more in curating their institutional presence, this effort does not translate into greater discoverability. Thus, while search rankings within categories do not appear systematically biased, the reduced number of visible links may still disadvantage female professors in terms of overall digital visibility.

Taken together, these results highlight how digital visibility is shaped by the interaction between algorithmic systems, individual presentation choices, and institutional infrastructure. They also reflect broader structural patterns: who appears where, how prominently, and through what types of content is not random; it is filtered through technical systems that rely on data structures, which may themselves encode or reflect gendered norms.

In light of our definition of algorithmic gender fairness, our findings suggest that current systems fall short of this ideal. Even when the intent may not be discriminatory, existing systems amplify disparities through uneven coverage, limited keyword matching, unclear ranking mechanisms, and visibility differences in general-purpose search results. These systems do not just reflect the real world—they actively reshape which parts of it are seen.

Fairness, therefore, cannot be evaluated purely by the absence of discriminatory intent or overt exclusion. It must also consider the cumulative effects of design decisions, platform constraints, and structural imbalances in source data. Gendered visibility gaps, even if subtle, are a form of representational inequality that algorithmic systems may unintentionally perpetuate.

\section{\uppercase{Conclusion and Future Work}}
In this paper, we introduced the concept of algorithmic gender fairness and evaluated it using heterogeneous data on German professors. By analysing gendered patterns in academic visibility across different institutional contexts and disciplines, we aimed to identify structural imbalances that may arise in algorithmic representations of expertise.

Our findings reveal nuanced but consistent gender differences in digital visibility. Search and retrieval algorithms do not exhibit overt forms of gender discrimination; however, subtle imbalances appear across various dimensions. Female professors were slightly more likely to complete their institutional profiles with CVs and publication lists, while male professors reported higher median numbers of publications. Yet, only a small number of self-reported publications could be retrieved from academic databases, highlighting mismatches between metadata, keyword representation, and retrieval mechanisms.

In Google search results, male professors were associated with a greater number of links overall, while female professors showed more variability, including more frequent cases of low link counts. Link categorisation and ranking further revealed gendered patterns: female professors’ links tended to appear in higher positions (i.e., closer to the top of the results list) in categories such as university websites, research profiles, and social media. Male professors' links, by contrast, were often ranked slightly lower (i.e., further down in the result list) in university websites and social media, but were more numerous overall. These differences likely reflect an interplay between platform algorithms, institutional curation, and self-presentation strategies.

While these patterns point to structural imbalances, they should be interpreted with caution. Factors such as outdated publication lists, common naming conventions, and differing levels of online activity likely contribute to the observed visibility gaps. The imbalances we observed are therefore not attributable to algorithmic bias alone, but emerge from the interaction of algorithmic processes with broader sociotechnical contexts.

Future research should expand on this foundation by incorporating more inclusive gender categories, extending the analysis beyond German academia, and examining additional disciplines. Integrating data from more publication databases and search engines would also allow for a broader assessment of visibility dynamics across digital ecosystems.

Longitudinal analyses and larger, more diverse datasets will be essential for disentangling the specific roles played by algorithmic systems, institutional infrastructures, and individual behaviours. In parallel, collaborative efforts involving academic institutions, search engine providers, and fairness researchers are needed to improve algorithmic transparency and accountability. Only by addressing both data and design can we move toward systems that fairly represent the diversity of academic expertise online.

\section*{\uppercase{AI Usage}}
The authors are not native English speakers; therefore, ChatGPT and Grammarly were used to assist with writing English in this work.

\section{\uppercase{Ethical Considerations}}
This paper did not involve direct interaction with human participants and relied solely on publicly available information found on university websites. As such, ethics approval from an institutional review board was not required. Personal data were collected manually with the intent to minimise misclassification, particularly in regard to gender inference. The student responsible for data collection was instructed to assign gender only when certainty was high and to otherwise mark entries as unknown. No personal or sensitive data beyond what was already publicly accessible were stored or analysed.

To protect the privacy and anonymity of the professors included in the dataset, we will not publish or share the collected data. We acknowledge the ethical limitations of inferring gender from names and pictures, and we explicitly address these limitations in the paper to promote transparency and encourage more inclusive data practices in future research.

\section{\uppercase{Adverse Impact Statement}}
This paper adopts a bias-preserving definition of algorithmic gender fairness, aiming to reflect real-world gender distributions without introducing or amplifying existing biases. While this approach supports transparency and alignment with observed data, it may also carry certain risks.

First, reflecting real-world distributions without intervention could be misused to justify existing gender inequalities, especially in contexts where structural bias is already present. Second, although we acknowledge the existence and importance of non-binary and gender-diverse identities, our empirical analysis is limited to binary gender categories due to data constraints. This limitation may contribute to the erasure of individuals who do not identify within the binary framework, especially if such approaches are widely adopted without critical adaptation. Finally, bias-preserving fairness may be misinterpreted as evidence of algorithmic neutrality, potentially obscuring the broader sociotechnical dynamics that shape inequality.

We therefore emphasise that fairness assessments should always be interpreted in light of context, data limitations, and the values underlying system design. We encourage future work to engage critically with fairness definitions and to explore approaches that address structural imbalances more directly.

\section*{\uppercase{Acknowledgements}}
This work was written by an author team working on different projects. Stefanie Urchs' project "Prof:inSicht" is promoted with funds from the Federal Ministry of Research, Technology and Space under the reference number 01FP21054. Matthias Aßenmacher is funded with funds from the Deutsche Forschungsgemeinschaft (DFG, German Research Foundation) as part of BERD@NFDI - grant number 460037581. Responsibility for the contents of this publication lies with the authors.

\bibliographystyle{apalike}
{\small
\bibliography{algo-gender-fairness}}

\end{document}